# Disentanglement of the electronic and lattice parts of the order parameter in a 1D Charge Density Wave system probed by femtosecond spectroscopy

H. Schäfer[1], V. V. Kabanov[2,3], M. Beyer[1], K. Biljakovic[4], and J. Demsar[1,2,3]

[1]*Physics Department and Center of Applied Photonics, Universität Konstanz, D-78457, Germany*
[2]*Zukunftskolleg, Universität Konstanz, D-78457, Germany*
[3]*Complex Matter Department, Jozef Stefan Institute, SI-1000, Ljubljana, Slovenia and*
[4]*Institute of Physics, Hr-10000 Zagreb, Croatia*

We report on the high resolution studies of the temperature (T) dependence of the $q = 0$ phonon spectrum in the quasi one-dimensional charge density wave (CDW) compound $K_{0.3}MoO_3$ utilizing time-resolved optical spectroscopy. Numerous modes that appear below $T_c$ show pronounced T-dependences of their amplitudes, frequencies and dampings. Utilizing the time-dependent Ginzburg-Landau theory we show that these modes result from linear coupling of the electronic part of the order parameter to the $2k_F$ phonons, while the (electronic) CDW amplitude mode is overdamped.

Femtosecond optical spectroscopy is becoming an important tool for investigation of the so called strongly correlated systems due to its intrinsic ability to determine the interaction strengths between various degrees of freedom which lead to fascinating phenomena like superconductivity or giant magnetoresistance. Low dimensional charge density wave (CDW) systems, with their inherently multi-component order parameter (modulation of carrier density is accompanied by the modulation of the underlying lattice) present no exception. In the past decade or so various one and two dimensional CDWs have been studied by time-resolved optical [1–6] as well as photoemission [7, 8] techniques. The initial focus of research was in identifying various components in the observed photoinduced transients with the corresponding ones obtained by standard time-averaging spectroscopic techniques, as well as in coherent control of the collective modes [9]. Recently it was shown that photoexcitation with an intense optical pulse can nonthermally drive the phase transition from the low temperature CDW state to a metastable state, characterized by a suppressed carrier modulation with the lattice remaining nearly frozen [6]. This observation has an important implication for the understanding of ultrafast relaxation processes in this class of materials, and, as we will show, for the general understanding of the cooperative phenomena leading to the appearance of the CDW state and the nature of their collective excitations.

In this Letter we present high resolution studies of T-dependent time-resolved reflectivity dynamics in a prototype quasi-1D CDW material $K_{0.3}MoO_3$. The high sensitivity achieved in this experiment enabled us to measure the T-evolution of the low frequency phonons with unprecedented resolution. We were able to show that not only the 1.68 THz (57 cm$^{-1}$) mode, that is commonly assigned to the collective amplitude mode (AM) of the CDW, shows softening upon increasing T towards $T_c = 183$ K, the phase transition to the normal metallic state. Qualitatively the same softening is observed also for a number of phonon modes that appear below $T_c$. The frequencies of these modes correspond well to the phonon frequencies at the $2k_F$ modulation vector as observed by neutron experiments [10, 11], and are shown to result from the linear coupling of the electronic part of the order parameter (EOP), $\Delta$, to $2k_F$ phonons. Utilizing the time-dependent Ginzburg-Landau (TDGL) model we were able to account for the T-dependence of mode frequencies, dampings and their amplitudes. Surprisingly, the coupling strengths to the EOP of all the modes that show softening is nearly the same. The analysis suggests that in $K_{0.3}MoO_3$ the non-adiabatic regime is realized, where electronic modulation does not adiabatically follow the lattice. The amplitude mode of the EOP, describing the initial recovery of the electronic density modulation, is shown to be an overdamped mode whose damping time diverges as $|\Delta|^{-2}$.

We studied the T-dependence of the PI reflectivity dynamics in single crystals of blue bronze $K_{0.3}MoO_3$ using an optical pump-probe technique. A commercial Ti:Sapphire amplifier producing 40 fs laser pulses at $\lambda = 800$ nm (photon energy of 1.55 eV) at a 250 kHz repetition rate was used as a source of both pump and probe pulse trains. The probe laser beam ($\vec{E}$) was polarized either along the chain direction ($\vec{b}$) or along the perpendicular (102) direction [12], while the pump beam was always polarized at an angle of 45° with respect to the probe polarization. The induced changes in reflectivity (R) were recorded utilizing a fast-scan technique, enabling high signal-to-noise levels. The excitation fluence was kept at 30 $\mu$J/cm$^2$: well below the non-linear regime [6], yet resulting in a large photoinduced reflectivity change, enabling the dynamic range of the signal of $\approx 10^4$.

Figure 1a) presents the induced reflectivity transients taken at 10 K with $\vec{E} \parallel \vec{b}$ (data for $\vec{E} \perp \vec{b}$ are shown in [13]). The transient can be decomposed into an electronic part, which shows a bi-exponential decay with timescales $\tau_1 \approx 0.2$ ps and $\tau_2 \approx 5$ ps [6], and a coherent part whose Fourier transform, obtained by the Fast Fourier Transform (FFT) analysis, is shown in panel b). The FFT shows numerous frequency components which can be attributed to the coherently excited phonon modes [1, 4, 6]. Most of the observed modes are seen also in Raman [14], however the far superior dynamic range and frequency resolution ($\approx 0.1$ cm$^{-1}$) of the data obtained by the time-resolved technique enables detailed study of their T-dependence.

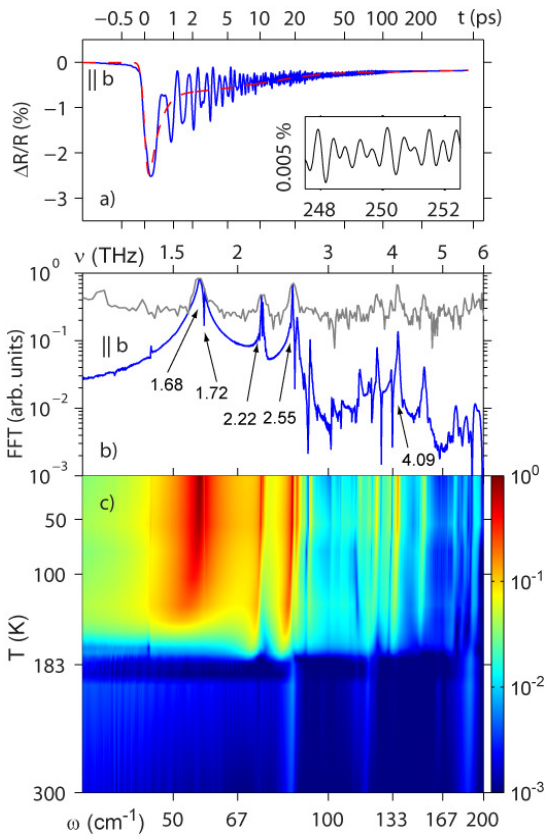

FIG. 1. (color online) a) Transient change in reflectivity of $K_{0.3}MoO_3$ with $\vec{E} \parallel \vec{b}$ at 10 K following photoexcitation with a 40 fs laser pulse. Red line is the fitted electronic transient. Inset: blow-up of the response near 250 ps. b) The FFT spectrum (amplitude) of the coherent parts of the signals (blue), compared to the recent Raman data (gray) from Ref.[14]. c) The T-dependence of the corresponding FFT spectrum in the range between 1-6 THz.

Figure 1 c) shows the T-dependence of the corresponding FFT spectrum. Most of the modes are seen only below $T_c$ implying that these modes result from the symmetry breaking in the CDW phase. Noteworthy, not only the 1.68 THz mode, which is commonly referred to as the amplitude mode of the CDW, but also numerous higher frequency modes show comparable softening as $T \to T_c$. To analyze the T-dependence of the modes we fit the FFT data with a sum of damped oscillators. To do so, we first fit the mode that has the strongest amplitude, subtract the resulting fit from the raw data, and perform the same routine on the residual signal to extract the data on the next most intense mode. The T-dependence of the 7 lowest frequency modes is shown in Fig. 2 b). We see that three modes in this frequency range show pronounced softening as $T \to T_c$, while the frequencies of four very narrow modes at 1.36, 1.72, 2.23 and 2.58 THz remain constant within the experimental accuracy. The frequencies of the most intense modes at 1.68, 2.22 and 2.55 THz, match well with the modes at $2k_F$ as seen in neutron experiments [10, 11] - see Fig. 2a). These observations suggest that these modes, observed in time-resolved experiments as well as in Raman (both probing at $q = 0$), originate from some type of "zone folding" mechanism, as argued earlier [4, 6, 14]. As we show below, this "folding" can be naturally explained by considering linear coupling of the EOP with the $2k_F$ phonons.

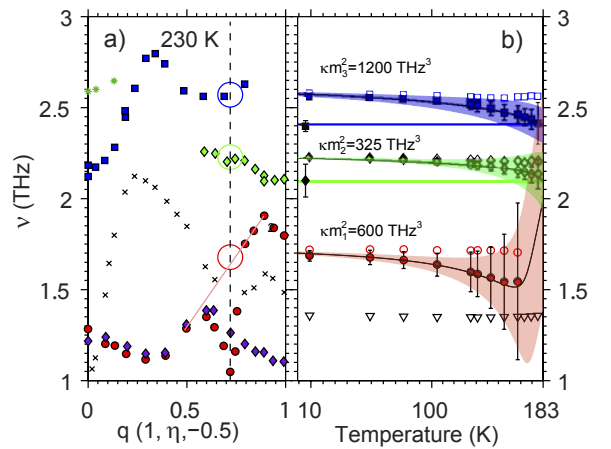

FIG. 2. (color online) a) Phonon dispersion in $(1,\eta,-0.5)$ direction at 230 K - reproduced from Ref.[10]. Dashed line corresponds to the CDW modulation wavevector, while the solid red line is the expected dispersion of the LA mode at $T \gg T_c$. b) The T-dependence of selected phonon modes: measured $\nu$ and $\Gamma/2\pi$ (solid symbols and bars) fit with the model (solid lines and shaded areas) - see also [13]. The values of $\kappa m_i^2$ (in units THz$^3$) obtained from the fit are shown. The frequencies (dampings) of the infrared modes at 6 K from Ref.[16] are shown by solid black symbols.

The CDW transition in $K_{0.3}MoO_3$ (space group $C2/m$), is characterized by the modulation wavevector $(1, \eta, -0.5)$ which corresponds to the vector $k_3$ of the Brillouin zone (BZ) in Kovalev's notation [15]. This vector is the non Lifshitz point of the BZ and describes the appearance of an incommensurate structural and electron density modulation; the modulation is incommensurate down to the lowest T with $\eta$ approaching 0.75 [12]. Vector $k_3$ belongs to the two-arms star of the wave vector with prongs transforming one to another via inversion ($K_{0.3}MoO_3$ retains the inversion symmetry in the CDW phase [16]). Therefore, the small group has two elements and two irreducible representations and the CDW phase transition is characterized by the two component order parameter. The order parameter can be represented by a complex number $\Delta = \Delta_1 + i\Delta_2$, where $\Delta_1$ and $\Delta_2$ are the real and imaginary parts, and $|\Delta| = (\Delta_1^2 + \Delta_2^2)^{1/2}$. Since below $T_c$ different modes appear at $q = 0$ we include in the thermodynamic potential the modes coupled linearly to the order parameter. Such a procedure does not make sense if one considers the thermodynamical properties of the system only. When discussing the dynamical properties of the system, such as the normal modes, these terms have to be included, as discussed in Ref.[17]. Therefore we define $\Delta_{1,2}$ as a purely EOP associated with the $2k_F$ carrier density modulation. It is in general linearly coupled with any lattice displacement at $2k_F$ which transforms as the $\tau_1$-representation of the small group. Deformations belonging to $4, 6, 8k_F$ may be coupled only to the higher orders of the EOP and are therefore not included. In the strictly incommensurate case, when the phase transition is associated with a non-Lifshitz wave-vector, the higher order invariants are also absent (the thermodynamic potential does not depend on the phase of the order parameter [18, 19]). The resulting thermodynamic potential can be written as

$$\phi = \phi_0 + \frac{1}{2}\alpha(T - T_{c0})(\Delta_1^2 + \Delta_2^2) + \frac{1}{4}\beta(\Delta_1^2 + \Delta_2^2)^2 \quad (1)$$
$$+ \frac{\Omega_0^2}{2}(\xi_1^2 + \xi_2^2) - m(\Delta_1\xi_1 + \Delta_2\xi_2).$$

Here $\phi_0$ corresponds to the high-T phase, $\alpha, \beta > 0$ are the standard GL constants, $\xi_{1,2}$ are generalized coordinates of displacements which transform as $\tau_1$ representations of the small group. $\Omega_0$ is the frequency of the vibrational mode at $T \gg T_{c0}$, where $T_{c0}$ is the bare critical temperature (in the absence of coupling to the lattice) and $m$ describes the strength of the coupling between the mode and the EOP. Here we assume that the effective mass of the mode is equal to 1. To minimize $\phi$ we choose the following (equilibrium) solution: $\xi_2^{(0)} = \Delta_2^{(0)} = 0$, $\Delta_1^{(0)} = \sqrt{\alpha(T_c - T)/\beta}$ and $\xi_1^{(0)} = \frac{m\Delta_1^{(0)}}{\Omega_0^2}$. The observable $T_c$ is renormalized due to the coupling with the displacement and is given by $T_c = T_{c0} + \frac{m^2}{\alpha\Omega_0^2}$. For the illustrative purpose we include only one mode coupled to the EOP, but the result can be generalized to all same symmetry modes at $2k_F$ that are coupled to $\Delta_{1,2}$ linearly [13].

Let us now consider the equations of motion of the EOP and the phonon mode assuming small fluctuations near their equilibrium positions, i.e. $\Delta_{1,2}(t) = \Delta_{1,2}^{(0)} + x_{1,2}(t)$ and $\xi_{1,2}(t) = \xi_{1,2}^{(0)} + y_{1,2}(t)$. We can assume the electronic mode to be overdamped, since the frequency of the bare mode $\omega = \sqrt{2}|\Delta|$ lies above the gap for single particle excitations, as in the case of a spin density wave [12]. The equations for the real parts of the order parameter $(x_1, y_1)$, describing the overdamped amplitude mode of the electronic channel and the Raman active lattice vibration, respectively, are

$$\dot{x}_1 = -2\kappa\alpha(T_c - T + \frac{m^2}{2\alpha\Omega_0^2})x_1 - \kappa m y_1 \quad (2)$$
$$\ddot{y}_1 = -\Omega_0^2 y_1 - mx_1,$$

$\kappa^{-1}$ being the analog of the friction coefficient. Similar analysis for the infrared modes [16, 20] is presented in [13]. The general solution of Eqs.(2) can be found in the form of $x_1 = a_1 \exp(\lambda_1 t)$ and $y_1 = b_1 \exp(\lambda_1 t)$ resulting in a cubic equation for $\lambda_1$:

$$\lambda_1^3 + 2\kappa\alpha(T_c - T + \frac{m^2}{2\alpha\Omega_0^2})\lambda_1^2 + \Omega_0^2\lambda_1 + 2\kappa\alpha\Omega_0^2(T_c - T) = 0. \quad (3)$$

At $T = T_c$ it follows that $\lambda_1^{(1)} = 0$, indicating that the relaxation time for the EOP diverges, while $\lambda_1^{(2,3)} = \pm i\Omega_0\sqrt{1 - (\frac{\kappa m^2}{2\Omega_0^3})^2} - \frac{\kappa m^2}{2\Omega_0^2}$. The mode frequency $\Omega(2\pi\nu) = \text{Im}(\lambda_1)$ and damping $\Gamma = \text{Re}(\lambda_1)$ at $T = T_c$ exactly correspond to the values of the matching infrared mode [13]. Indeed, for both 2.22 and 2.55 THz modes, we find that the mode frequencies near $T_c$ match well the frequencies of the infrared modes, discussed in Ref. [16] using a similar scenario [16, 20].

The above model describes both adiabatic and non-adiabatic limits. In the adiabatic limit, when $\Gamma_{EOP} \gg \Omega_0$, the phonon mode is a true soft mode with $\Omega_0 \to 0$ at $T_c$ - see [13]. In the intermediate (non-adiabatic) case, however, $\Omega$ shows softening only until $\Gamma_{EOP} \approx \Omega_0$, while in the vicinity of the phase transition $\Omega$ can also increase. The solutions of Eq.(3) for the intermediate case are shown in insert to Fig. 3a). Indeed, in the entire class of materials $\Gamma_{EOP} \approx \Omega_0$, with damping of the fast electronic component showing critical slowing down towards $T_c$ [1, 3, 5] - see Fig. 3a. This suggests, that the fast electronic decay process can be identified as the overdamped mode of the EOP.

Using this simple model we were able to fit the T-dependence of $\Omega$ and $\Gamma$ for the 3 most intense phonon modes - see Fig. 2b). The agreement between the measured frequencies (symbols) and $\Omega$'s from the model (solid lines) is nearly perfect. A very good agreement is also found between the measured (error bars) and model (shaded areas) damping constants. Given the simplicity of the model, assuming a GL T-dependence of $|\Delta|^2$ over the entire T-range, the fact that other processes (e.g. dephasing) can also contribute to the damping, the agreement is excellent. It is noteworthy that the coupling strengths $m_i$ between the EOP and the three most intense phonons (see Eqs.(1-3)), obtained by fitting to the model, are nearly identical - see Fig. 2b).

Fig. 3b) shows the T-dependence of mode amplitudes, normalized to their low temperature value. All modes that show pronounced softening near $T_c$ show very similar T-dependence of amplitude, which can be well ex-plained within our model. Assuming that the order parameter is small we can expand the dielectric constant near the CDW phase transition in powers of the order parameter $\Delta_{1,2}$ and variables $\xi_{1,2}$ [21]:

$$\epsilon(k=0) = \epsilon_0 + c_1(\Delta_1^2 + \Delta_2^2) + c_2(\Delta_1\xi_1 + \Delta_2\xi_2) + ... \quad (4)$$

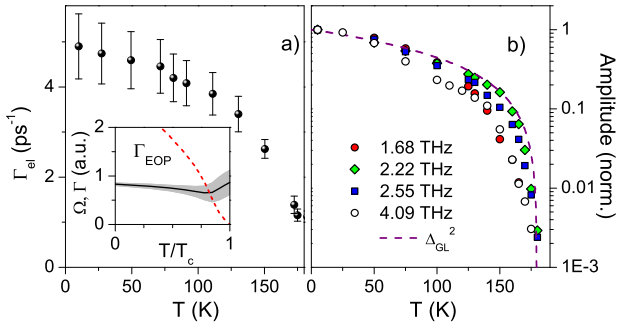

FIG. 3. (color online) a) The measured T-dependence of the fast electronic damping showing critical slowing down at $T_c$. Insert shows the model simulation with one mode linearly coupled to the EOP, where the solid line (shaded area) represent the phonon mode frequency (damping), while the dashed line represents the damping of the overdamped EOP. b) The T-dependence of normalized amplitudes of modes that show softening. It follows well the expected $|\Delta|^2$ T-dependence (dashed line).

Here $\epsilon_0$ is the dielectric constant of the high-T symmetric phase, and $c_1$ and $c_2$ are real constants (linear terms in the expansion are not allowed by symmetry). It follows that the induced change in the displacement ($\xi_{1,2}$) should result in a change in $\epsilon$ (or $R$) that is proportional to $\Delta_{1,2}$. However, since the change in the displacement is itself proportional to the EOP, the mode amplitude - as determined by the time-resolved experiment - should follow the $|\Delta|^2$ T-dependence. The data are indeed well fit to the GL T-dependence of $|\Delta|^2_{GL} \propto (T_c - T)/T$. Amplitudes of the weak satellite modes, which show nearly no T-dependence of frequency, are however found to show a substantially faster decrease with T [13]. This suggests that the satellite modes are either $q = 0$ modes, which are amplified below $T_c$ due to coupling to the neighboring "folded" modes (like 2.23 and 2.58 THz modes), or are the result of a higher order coupling to the EOP (1.72 and 1.36 THz modes) [13].

In conclusion, high resolution T-dependence studies of photoinduced reflectivity changes in the quasi 1D CDW compound $K_{0.3}MoO_3$ enabled us to track the T-dependence of the coherently generated phonon modes with unprecedented sensitivity. Numerous modes that appear below $T_c$, and show comparable softening as $T_c$ is approached, are observed. By applying TDGL analysis we were able to show, that these modes are a result of the linear coupling of the EOP to the $2k_F$ phonons where the EOP *does not* adiabatically follow the lattice modulation. This interpretation presents an alternative to the fluctuation scenario [10, 22], answering the longstanding question why no phonon shows full softening near $T_c$ in this class of materials. The fact that a fast electronic dynamics, whose decay time diverges near $T_c$, is observed in many CDW compounds [1, 3, 5] lead us to identify this mode as the overdamped (electronic) amplitude mode. The disentanglement of the electronic and lattice parts of the order parameter on the very short time-scale, demonstrated here for the CDW systems, could be however operational in a broader class of materials undergoing structural phase transitions.

We wish to acknowledge Stefan Eggert for his help with various aspects of data acquisition, D. Sagar for providing us the low temperature Raman data, and valuable discussions with V. Pomjakushin, J. P. Pouget, J. E. Lorenzo, L. Degiorgi, and P. Monceau. The work was supported by the Sofja-Kovalevskaja Grant from the Alexander von Humboldt Foundation, Zukunftskolleg and CAP at the University of Konstanz.

[1] J. Demsar, K. Biljakovic, D. Mihailovic, *Phys. Rev. Lett.* **83**, 800 (1999).
[2] J. Demsar, et al., *Phys. Rev. B* **66**, 041101 (2002).
[3] K. Shimatake, Y. Toda, and S. Tanda, *Phys. Rev. B* **75**, 115120 (2007).
[4] D.M. Sagar et al., *J.Phys. Cond. Mat.* **19**, 436208 (2007).
[5] R.V. Yusupov, et al., *Phys. Rev. Lett.* **101**, 246402 (2008).
[6] A. Tomeljak, et al., *Phys. Rev. Lett.* **102**, 066404 (2009).
[7] L. Perfetti, et al., *Phys. Rev. Lett.* **97**, 067402 (2006); L. Perfetti, et al., *New J. of Phys.* **10**, 053019 (2008).
[8] F. Schmitt, et al., *Science* **321**, 1649 (2008).
[9] D. Mihailovic, et al., *Appl. Phys. Lett.* **80**, 871 (2002); T. Onozaki, Y. Toda, S. Tanda, R. Morita, *Jap. J. of Appl. Phys.* **46**, 870 (2007).
[10] J.P. Pouget, et al., *Phys. Rev. B* **43**, 8421 (1991).
[11] C. Escribe-Filippini, J.P. Pouget, R. Currat, B. Hennion, J. Marcus, in Lecture Notes in Physiccs **217** (Springer Verlag, 1985), p. 71-75.
[12] G. Grüner, Density waves in Solids, Addison-Wesley, 1994.
[13] Supplementary online material.
[14] D.M. Sagar, et al., *New J. of Phys.* **10**, 023043 (2008).
[15] O.V. Kovalev, Irreducible representations of the space groups. Gordon and Breach, 1965.
[16] L. Degiorgi, B. Alavi, G. Mihály, G. Grüner, *Phys. Rev. B* **44**, 7808 (1991).
[17] D.G. Sannikov, *Phys. of Solid Stat* **50**, 746 (2008).
[18] Y.A. Izyumov and V.N. Syromyatnikov, Fazovye Perekhody I Simmetriya Kristallov, Nauka, Moskow (1984) (in Russian).
[19] J. Ollivier, et al., *Phys. Rev. Lett.* **81**, 3667 (1998).
[20] M.J. Rice, *Phys. Rev. Lett.* **37**, 36 (1976).
[21] V.L. Ginzburg, Sov. Phys. Usp. 5, 649 (1963).
[22] E. Tutis, S. Barisic, Phys. Rev. B. 43, 8431 (1991).

Supplementary material to *"Disentanglement of the electronic and lattice parts of the order parameter in a 1D Charge Density Wave system probed by femtosecond spectroscopy"*

H. Schäfer[1], V. V. Kabanov[2,3], M. Beyer[1], K. Biljakovic[4], and J. Demsar[1,2,3]

[1]Physics Department and Center of Applied Photonics, Universität
Konstanz, D-78457, Germany

[2]Zukunftskolleg, Universität Konstanz, D-78457, Germany

[3]Complex Matter Department, Jozef Stefan Institute, SI-1000, Ljubljana, Slovenia

[4]Institute of Physics, Hr-10000 Zagreb, Croatia

**Contents**

1. Experimental Set-up
2. The Raman and IR active q=0 modes in the CDW state obtained within the TDGL model
3. Extending the model to multiple phonons: numerical analysis
4. Phonon spectra probed with different probe polarizations with respect to crystal axes
5. The temperature dependence of the phonon parameters
6. On the possible nature of the narrow satellite modes

## 1. Experimental Set-up

The data were recorded using the experimental set-up, whose simplified scheme is shown in Fig. 1. The incoming beam (800 nm, 50 fs, 250 kHz) is split into a pump and a probe arm using a beam splitter. We utilize the so called fast scan technique, where the time-delay between the pump and the probe pulse is delayed via a mechanical shaker. The shaker operates at a frequency of about 20 Hz, and its maximum displacement corresponds to the time delay of 100 ps. To achieve larger time delays, as in the present experiment, we use an additional computer controlled delay stage in the pump arm, enabling a maximum time delay of 4 ns. We use the differential detection scheme (the signal from the beam reflected from the sample's surface is subtracted from the reference beam) and averaging on the oscilloscope to achieve high signal-to-noise ratio. The pump beam has been focused down to a 100 μm diameter spot with a 300 mm lens, while the probe beam is focused with a 150 mm lens to a spot with a 50 µm diameter to ensure probing of a homogenously excited sample area.

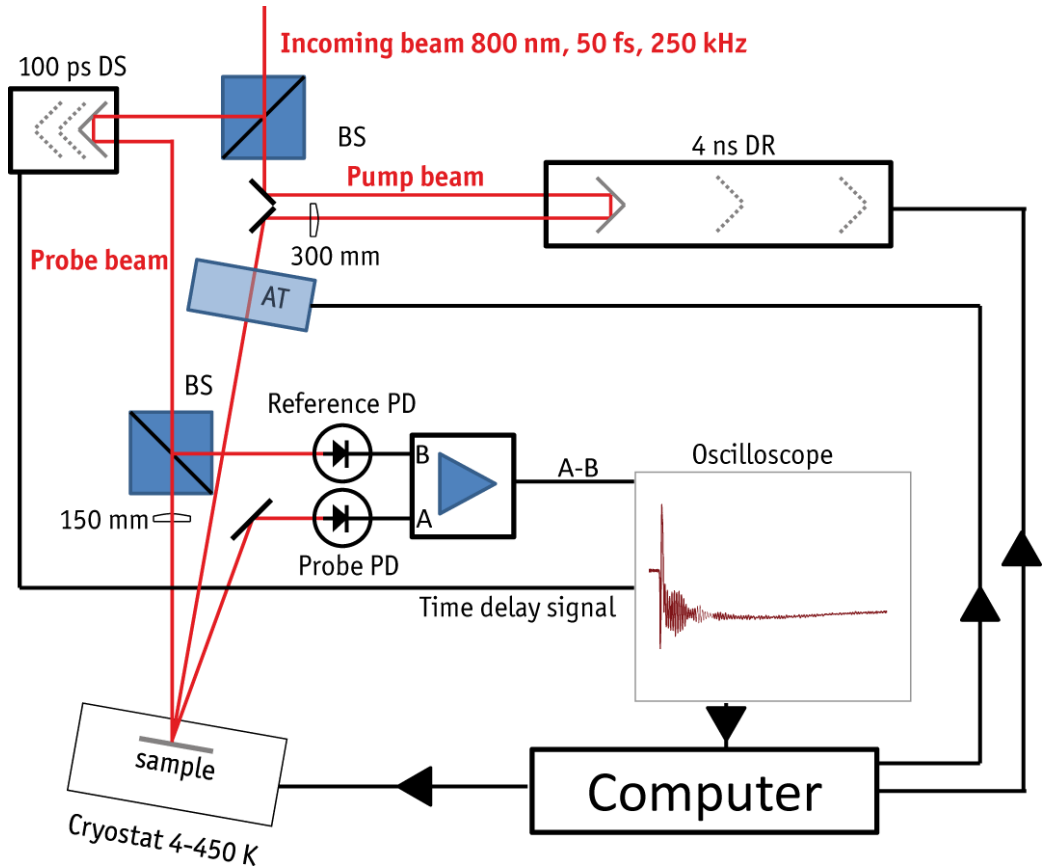

Fig. 1: Simplified scheme of the fully automatic setup that was used to retrieve the time resolved pump probe data. BS: beam splitter, PD: photo diode, DS: delay shaker, DR: delay stage, AT: attenuator, A: probe signal, B: reference signal.

## 2. Raman and IR active q=0 modes in the CDW state obtained within the TDGL model

The main motivations to apply the TDGL analysis in different limiting cases (adiabatic to non-adiabatic) are the observation of an overdamped (electronic) mode, whose damping time shows critical slowing down at $T_c$, the incomplete softening of several low frequency modes, as well as the peculiar temperature dependence of the damping of the phonon modes. As far as an overdamped electronic mode is concerned, we can easily rule out other alternative explanations. If this component was due to simple electron-phonon thermalization [1], no temperature dependence should be observed in contrast to experimental results. Alternatively, if this was an overdamped phonon mode, the damping should increase with temperature and not decrease as observed.

As has been shown in the main text we can write the free energy as

(1) $$\varphi = \varphi_0 + \frac{1}{2}\alpha(T-T_{c0})(\Delta_1^2+\Delta_2^2) + \frac{1}{4}\beta(\Delta_1^2+\Delta_2^2)^2 + \frac{\Omega_0^2}{2}(\xi_1^2+\xi_2^2) - m(\Delta_1\xi_1+\Delta_2\xi_2).$$

Assuming small fluctuations near their equilibrium positions, the equations of motion of the EOP and the phonon mode are derived. The equations for the real parts of the order parameter ($x_1$,$y_1$), describing the overdamped amplitude mode of the electronic channel and the Raman active (symmetric) lattice vibration, respectively, are:

(2) $$\dot{x}_1 = -2\kappa\alpha(T_c + \frac{m^2}{2\alpha\Omega_0^2} - T)x_1 - \kappa m y_1$$

$$\ddot{y}_1 = -\Omega_0^2 y_1 - m x_1$$

The corresponding equations of motion for the two infrared active modes are:

(3) $$\dot{x}_2 = -\kappa\frac{m^2}{\Omega_0^2}x_2 - \kappa m y_2$$

$$\ddot{y}_2 = -\Omega_0^2 y_2 - m x_2$$

Equations (2),(3) describe the overdamped electronic relaxation channel and Raman and infrared (IR) active lattice vibrations, respectively. Clearly, each phonon mode at $\pm 2k_F$ (in the high temperature phase) generates in the low temperature phase one Raman active and one infrared active mode with the frequencies close to $\omega(2k_f)$ in agreement with [2]. Moreover, comparison of the equations (2) and (3) directly reveals that at $T_c$ the IR mode frequency and damping are equal to that of the Raman active mode. Unlike for the case of the Raman active mode, the IR active mode shows no temperature dependence in the first approximation (see Eq.(3)). One should keep in mind, however, that as soon as there is some coupling between the IR and Raman modes (e.g. via higher order gradient terms of the order parameter in the thermodynamic potential) equations (2) and (3) become coupled.

Frequencies and dampings of the Raman active modes are determined by the equation:

(4) $$\lambda^3 + 2\kappa\alpha(T_c - T + \frac{m^2}{2\alpha\Omega_0^2})\lambda^2 + \Omega_0^2\lambda + 2\kappa\alpha\Omega_0^2(T_c - T) = 0$$

It is interesting to recover the standard amplitude and phase modes from this equation. If we assume that the electronic part is fast and adiabatically follows the lattice coordinates, i.e. $\kappa \to \infty$, we obtain the standard result for the temperature dependence of the amplitude mode as $\Omega \to 0$ as $T \to T_c$. With $\kappa \to \infty$, we can neglect the cubic and linear terms in (4) and obtain:

$$(5) \quad \Omega = \Omega_0 \sqrt{\frac{T_c - T}{T_c + \frac{m^2}{2\alpha\Omega_0^2} - T}}$$

Similarly, for $\kappa \to \infty$ and putting T=T$_c$ (which transforms Eq. (4) to the equation for the infrared mode) leads to the known result that the phase mode is gapless with 0 frequency at *k=0* [3].

### 2.1. The temperature dependence of the mode frequency and damping

The most interesting parameter in our context is temperature. As follows from Eq. (4) at $T = T_c$ $\lambda_1^{(1)} = 0$ indicating that the relaxation time for the overdamped electronic mode diverges. On the other hand, the phonon frequency shows softening in most of the temperature range, while in the close vicinity of $T_c$ it increases towards $T_c$.

The theory is in general applicable both in adiabatic and non-adiabatic limits (in the adiabatic limit the electron system adiabatically follows the lattice motion, while in the extreme non-adiabatic limit the electronic damping becomes slower than the inverse phonon frequency). However, in the close vicinity of $T_c$ the model is expected to fail, since in this case fluctuation, which are not included in the model, become important. Formally, the validity of the Ginzburg-Landau theory is limited to the temperature range:

$$(6) \quad Gi \ll \left|\frac{T - T_c}{T_c}\right| \ll 1,$$

where *Gi* is the Ginzburg number [4]. *Gi* is denoting the range where fluctuation of the order parameter are large, i.e. when $|\langle\Delta\rangle| < \sqrt{\langle\Delta^2\rangle - \langle\Delta\rangle^2}$. Therefore, the solutions plotted in Figs. 2 and 3 become invalid in the close vicinity of T$_c$.

In Fig. 2 the temperature dependence of the damping of the electronic mode and frequency and damping of the linearly coupled phonon mode is shown for different values of the bare electronic damping.

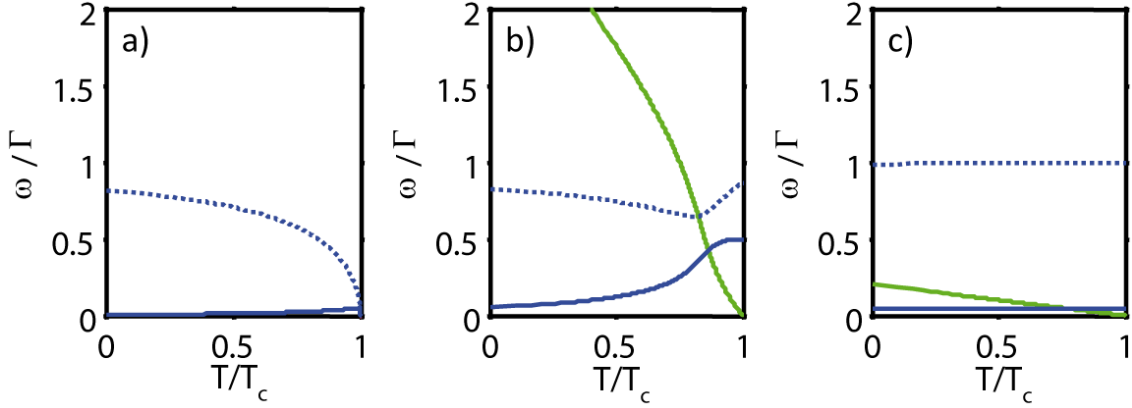

Fig. 2: The temperature dependence of the solution (green line: damping of the electronic mode, dotted blue line: phonon mode frequency, and solid blue line damping of the phonon mode). These are shown for different limiting cases: a) Adiabatic scenario: Strong electronic damping ($\Gamma_e=10\cdot\Omega_0$ – not shown, out of scale) leads to true soft mode behavior. b) Intermediate regime: Electronic damping of the order of the mode frequency ($\Gamma_e=\Omega_0$) leads to mode softening at lower temperatures. Near $T_c$, when electronic damping becomes too slow (near $T_c$) the mode hardens again. c) Fully non-adiabatic case ($\Gamma_e=0.1\Omega_0$): Only very weak hardening of the phonon mode is observed.

## 3. Extending the model to multiple phonons: numerical analysis

Starting from equation (1) we can extend this model to more than only one phonon mode by introducing a sum over all phonon modes $i$. We are considering only the symmetric Raman modes (IR modes are decoupled from Raman modes as shown above). The free energy is then given by:

(7) $$\phi = \phi_0 + \frac{\alpha}{2}(T-T_{c0})\Delta^2 + \frac{\beta}{4}\Delta^4 + \sum_i \frac{\Omega_i^2}{2}\xi_i^2 - m_i\Delta\xi_i .$$

Here $m_i$ are the coupling constants of the modes, while $\Omega_i$ are the mode frequencies and $\xi_i$ the mode amplitudes. (Note that in this case the index $i$ denotes the mode quantum number.)

This leads to a system of linear differential equations similar to Equation (2). To solve the system of linear differential equations numerically we reduce the system to the system of first order differential equations by introducing dummy variables. E.g. Equation (2) is rewritten as ($\psi$ is the dummy variable for the derivative of y):

(8)
$$\dot{x} = -2\kappa\alpha\left(T_c + \frac{m^2}{\alpha\Omega_0^2} - T\right)x - \kappa m y$$
$$\dot{y} = \psi$$
$$\dot{\psi} = -\Omega_0^2 y - mx$$

Searching for the solution of Eqs.(8) in the form $v = \tilde{v}_0 e^{\lambda t}$, where $\tilde{v}_0$ is the complex amplitude of the variables $x$, $y$, $\psi$, we obtain the eigenvalue equation

(9)
$$\lambda\tilde{x}_0 = -2\kappa\alpha\left(T_c + \frac{m^2}{\alpha\Omega_0^2} - T\right)\tilde{x}_0 - \kappa m\tilde{y}_0$$
$$\lambda\tilde{y}_0 = \tilde{\psi}_0$$
$$\lambda\tilde{\psi}_0 = -\Omega_0^2\tilde{y}_0 - m\tilde{x}_0$$

The eigenfrequencies and dampings may be calculated numerically as eigenvalues of the matrix:

(10)
$$\begin{pmatrix} -2\kappa\alpha\left(T_c + \frac{m^2}{\alpha\Omega_0^2} - T\right) & -\kappa m & 0 \\ 0 & 0 & 1 \\ -m & -\Omega_0^2 & 0 \end{pmatrix}.$$

Eq. (9) corresponds to the previous case described by the cubic equation - Eq. (4). For multiple phonons the system of equations of motion is written as:

(11)
$$\dot{x} = -2\kappa\alpha\left(T_c + \frac{m_1^2}{\alpha\Omega_1^2} + \frac{m_2^2}{\alpha\Omega_2^2}... - T\right)x - \kappa m_1 y_1 - \kappa m_2 y_2...$$
$$\dot{y}_1 = \psi_1$$
$$\dot{\psi}_1 = -\Omega_1^2 y_1^1 - m_1 x$$
$$\dot{y}_1^2 = \psi_1^2$$
$$\dot{\psi}_2 = -\Omega_2^2 y_2 - m_2 x$$
....

where the indices represent the phonon mode quantum numbers.
Transforming Eqs. (11) to the matrix form similar to Eqs. (9),(10) we obtain the eigenvalue equation of 1+2N order, where N is the number of phonon modes coupled to the EOP. For N>1 this equation is analyzed numerically. Even for N=1 the solution of the eigenvalue equation away from $T_c$ is not straight forward.

In Fig. 3) we present an example of two phonon modes coupled linearly to the EOP. The coupling strength is the same for both modes, but the bare phonon frequencies are different for the three subplots. All plots shown are in the intermediate regime where ($\Gamma_e \approx \Omega_0$).

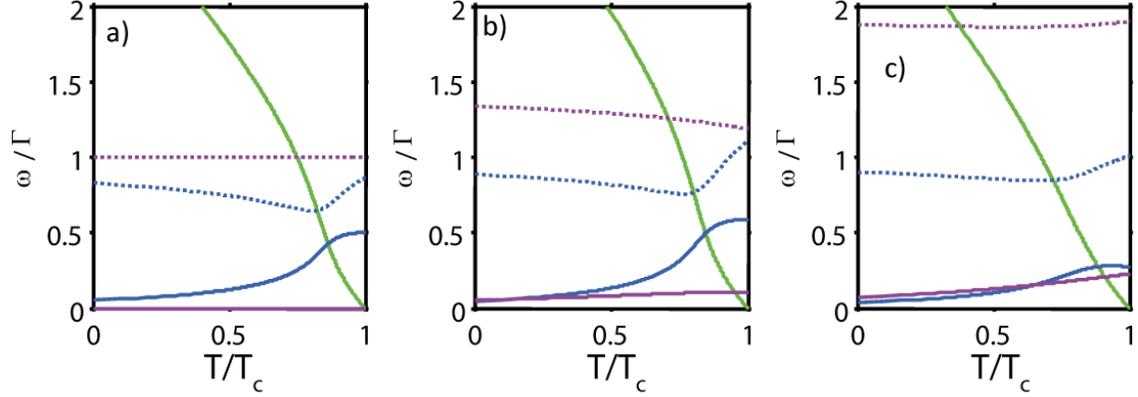

Fig. 3: Two phonons coupled linearly to the electronic part of the order parameter. a) Two modes with the same frequency ($\omega_1=\omega_2=1$). b) Two modes with frequencies ($\omega_1=1$ and $\omega_2=1.5$). c) Two modes with frequencies ($\omega_1=1$ and $\omega_2=2$). (green line: damping of the electronic mode, dotted blue/violet line: phonon modes frequencies, and solid blue/violet lines damping of the phonon modes).

Panel a) of Fig. 3) shows two modes having the same bare frequency. As expected, one of the modes shows no temperature dependence of frequency and damping. (This is also true for more than two modes with the same frequency). In panel b) the ratio of the bare frequencies is 1.5. This provides two features: The high frequency mode shows softening throughout the entire temperature range, while the low frequency mode experiences an upturn which overcomes its bare frequency near $T_c$. If the modes are far separated from each other, as shown in panel c), interaction becomes weaker and the temperature dependence become less pronounced.

## 4. Phonon spectra probed with different probe polarizations with respect to the crystal axes

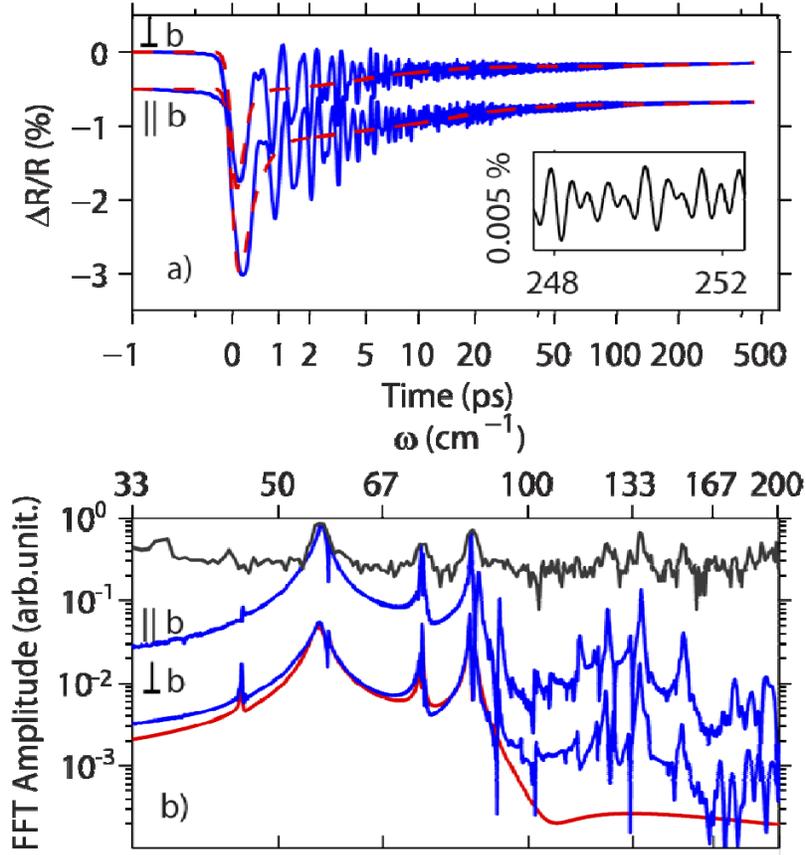

Fig.4: a) Comparison of the photoinduced reflectivity traces recorded at 10 K for both polarizations of the probe beam with respect to the crystal axes. b) The corresponding FFT spectra (amplitude). The red line is the fit accounting for modes at 1.36 THz (45.3 cm$^{-1}$), 1.68 THz ( 56 cm$^{-1}$), 1.72 THz ( 57.3 cm$^{-1}$), 2.21 THz ( 73.6 cm$^{-1}$), 2.23 THz ( 74.3cm$^{-1}$), 2.55 THz (85 cm$^{-1}$) and 2.58 THz ( 86 cm$^{-1}$).

Fig. 4 presents the comparison of the photoinduced reflectivity traces, recorded for both polarizations of the probe beam with respect to crystal axes, and their FFT (amplitude) spectra. We used a Lorentzian function to extract the characteristic parameters of the phonon modes:

$$(12) \quad \tilde{x}(\omega) = \frac{\tilde{x}_0}{\sqrt{(\omega_0 - \omega)^2 + \Gamma^2}}$$

where $\tilde{x}_0$ is the complex amplitude, $\omega_0$ the center frequency and $\Gamma$ the phonon damping.

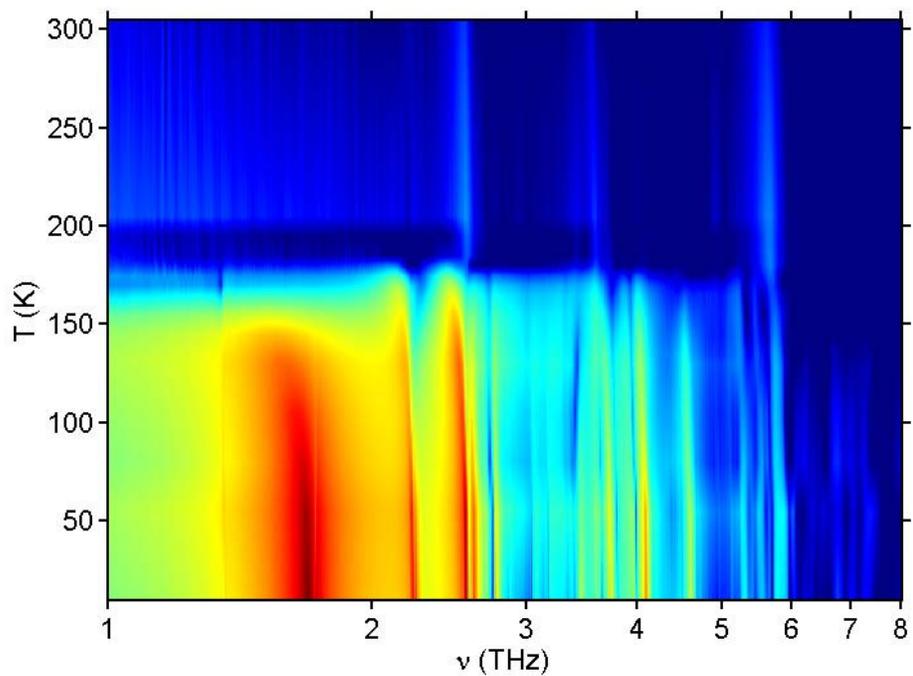

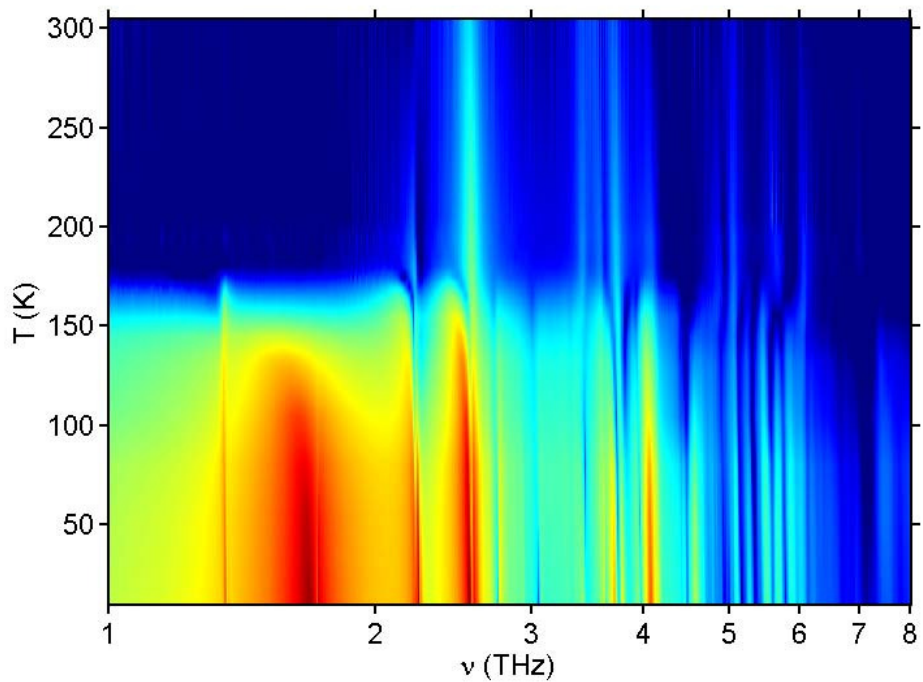

Fig. 5: The temperature dependent phonon spectrum probed with light polarized along the chain direction (b-axis) - top panel, and with probe light polarized perpendicularly to chain direction - bottom panel.

In both spectra (see Fig. 4 and Fig. 5 for their temperature dependencies) the main modes are at the same frequency and show the same temperature dependence. The weak satellite modes, which show almost no temperature dependence of their frequencies, are in general more pronounced in the configuration, where the probe polarization is perpendicular to chain direction. Moreover, the satellite modes at 2.58 THz as well as at 2.23 THz, are found to persist up to high temperatures, suggesting that they are the q=0 modes of the high temperature phase.

## 5. The temperature dependence of the phonon parameters

As discussed in the main text, the mode parameters were obtained by fitting the FFT data with the sum of Lorentzian line fits. Each line is determined by the four parameters: Amplitude, frequency, damping and phase. The amplitude, frequency and damping have been discussed in the manuscript. However, the damping was shown on the linear scale (given by bars in Fig. 2 of the main text). In Fig. 6 a) we show the temperature dependence of the corresponding decay times on the semi-log plot. Importantly, the temperature dependencies of the dampings of the modes cannot be accounted for by a standard anharmonic decay model - see dashed curves in Fig. 6 b), where for a phonon of energy $\hbar\omega_0$ the T-dependence of damping is given by $\Gamma(T) = \Gamma(0)\left(1 + 2/(e^{\hbar\omega_0/2k_BT} - 1)\right)$ [5]. On the other hand $\Gamma(T)$ is well reproduced by our model – see solid lines in Fig. 6 b) – where single parameter (the coupling strength between the $2k_F$ phonon mode and the EOP) is used to describe both T-dependence of softening and damping (see Fig. 2 of the main text).

For completeness, we are showing in Fig. 7 also the T-dependence of the phases of the coherently excited modes. As was discussed by Stevens et al. [6], the coherent phonon generation can be described by the two stimulated Raman tensors. This model contains both, the so called "impulsive" [6,7] and "displacive" [8] limiting cases, where the phase in the pure "impulsive" limit is $\pi/2$, while in the pure "displacive" limit it is 0. At low temperatures, the main modes show cosine like dynamics (as in the "displacive" limit), where phase is close to 0. The phases change as the critical temperature is approached, in a quite dramatic way.

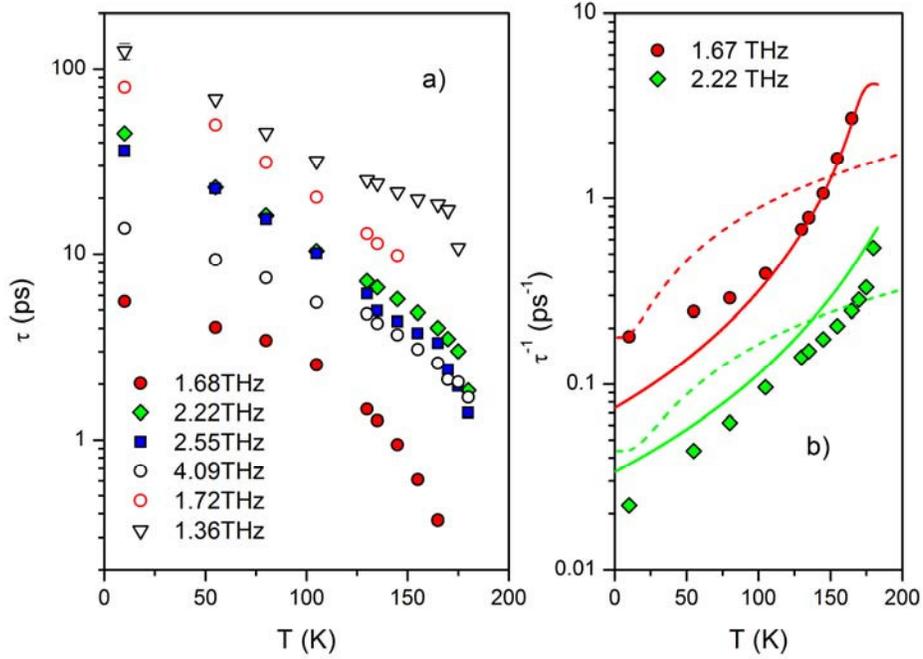

Fig. 6: a) The T-dependence of the decay times of the modes discussed in the main text. The error bars extracted from the fit are typically smaller than the symbol size, except for 1.36 THz mode at 10 K. b) The T-dependence of damping for two modes (1.68 and 2.22 THz) compared to the fit with the model (solid lines) and the standard anharmonic decay model (dashed lines).

For the general case (including both displacive and impulsive limits) the phase of the coherent phonon can be evaluated [9]. To be able to apply this model, however, one would need the data on the temperature dependence of the derivative of the real part of the dielectric function with respect to frequency at the optical frequency, which do not exist. Secondly, and more importantly, the modes that we are discussing are linearly (or even higher order - see Section 6) coupled to each other via the electronic part of the order parameter. It is most probably that the coupling between the modes, together with the strong temperature dependence of the dielectric function, and all the damping times, give such a dramatic temperature dependence of the phases. Given the fact, that these questions are far beyond the central topic of this paper, we are leaving this issue for future publication.

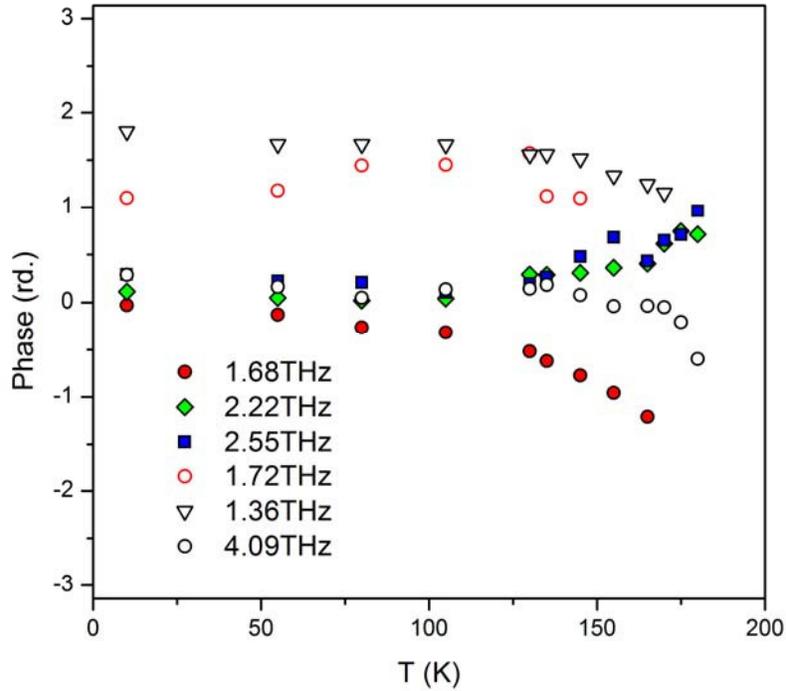

Fig. 7: The temperature dependency of the phase of the phonon modes, where each mode (*i*) follows $A_i * \exp(-t/\tau_i) * \cos(\omega_i t - \phi_i)$, where t is the time delay after photoexcitation. The error-bars for all the data points are about 0.2 rd., determined by the uncertainty of the zero time delay of about 20 fs.

## 6. On the possible nature of the narrow satellite modes

We should discuss the possible nature of narrow satellite modes which are located in the close vicinity of the main T-dependent modes and show no measurable T-dependence of frequency. As discussed above, the modes at 2.23 and 2.58 THz are, though very weak, observed also above $T_c$ and can be therefore attributed to q=0 modes of the high temperature phase. The mode at 1.72 THz is, despite the extremely high sensitivity, not resolved above ≈ 150 K.

The modes that are present only below $T_c$ could in principle arise due to higher order coupling to the electronic part of the order parameter; i.e. they could correspond to $4k_F$, $6k_F$ … phonons of the high temperature phase. Because of the higher order coupling to the electronic part of the order parameter one would expect that: a) their frequency shows much weaker temperature dependence and b) that their amplitude would be proportional to higher powers of the order parameter, $|\Delta|^4, |\Delta|^6$ … This explanation could well describe the mode at 1.72 THz.

The modes at 2.23 and 2.58 THz are observed also above $T_c$, suggesting that they are q=0 phonons of the high temperature phase. For the q=0 phonon, whose frequency lies in close proximity to the "folded" mode (the mode that results due to linear coupling to the electronic part of the order parameter), coupling to the folded mode is expected. This coupling could result in a substantial increase in the q=0 phonon intensity below $T_c$, as experimentally observed (See Fig. 5 for the 2.23 THz mode). In such a case, terms $(a_i + b_i|\Delta|^2)\zeta_i$ should be included in the expansion of the dielectric function (Eq.(4) of the main text), where $\zeta_i$ is the generalized phonon coordinate while $a_i$ and $b_i$ are constants ($a_i$ is in general temperature dependent). Using the same argument as for the "folded" modes the resulting q=0 phonon amplitudes should show $(a_i + b_i|\Delta|^2)^2$ T-dependence. Indeed, as shown in Figure 8, amplitudes of 1.72 and 2.23 THz modes decreases much faster than for the case of "folded" modes. Red line presents the fit with $(a_1 + b_1|\Delta|^2)^2$, where $a_1$ was assumed to be T-independent. The fit shows a good agreement with the data.

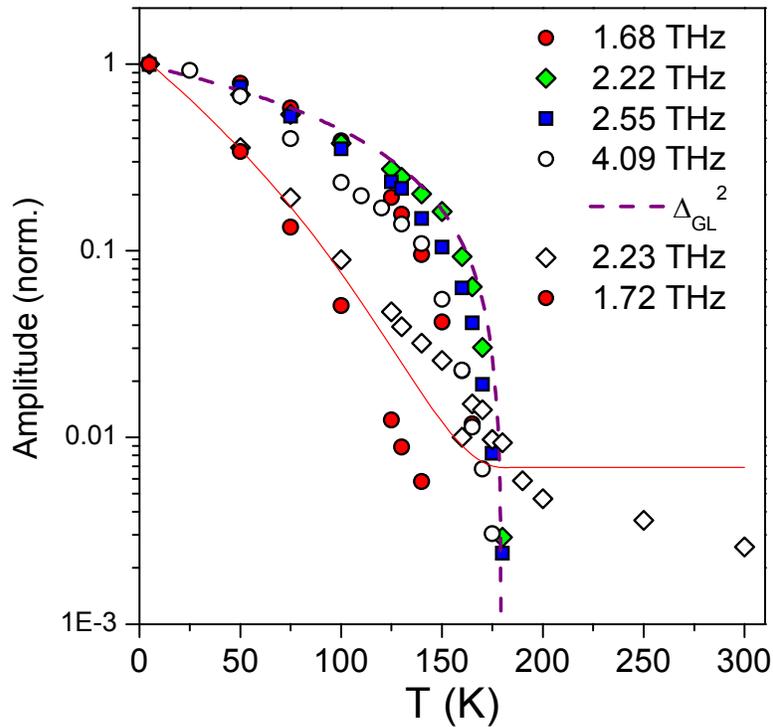

Fig. 8: The temperature dependence of the amplitudes of several phonon modes (normalized to their lowest temperature value). While the main modes follow the $|\Delta|^2$ temperature dependence, the amplitude of satellite modes decreases faster (see e.g. modes at 1.72 and 2.23 THz). The one for 2.23 THz mode can be well fit with $(a_1 + b_1|\Delta|^2)^2$ - red line (here $a_1$ was assumed to be temperature independent).

**References**

[1] see. e.g. R. H. M. Groeneveld, R. Sprik, and A. Lagendijk, Phys. Rev. B 51, 11433, (1991) and references therein.

[2] M.J. Rice, Phys. Rev. Lett., 36 (1976).

[3] G. Grüner, Density waves in Solids, Addison-Wesley, 1994.

[4] G. Klemens, Phys. Rev. 148, 845 (1966); M. Balkanski, R. F. Wallis and E. Haro, Phys. Rev. B 28, 1928 (1993).

[5] V. L. Ginzburg, Fizika Tverdogo Tela 2, 2031 (1960); Sov. Phys. Solid State 2, 1824 (1961).

[6] T. E. Stevens, J. Kuhl, and R. Merlin, Phys. Rev. B 65, 144304 (2002).

[7] L. Dhar, J.A. Rogers, K.A. Nelson, Chem. Rev. 94, 157 (1994);

[8] H.J. Zeiger et al., Phys. Rev. B 45, 768 (1992).

[9] D. M. Riffe and A. J. Sabbah, Phys. Rev. B 76, 085207 (2007)